\documentclass[aps,prb,twocolumn,showpacs]{revtex4}

\usepackage{graphicx}
\usepackage{amsmath}
\usepackage{amssymb}

\begin{document}

\title{Anisotropic dressing of electrons in electron-doped cuprate superconductors}


\author{Shuning Tan, Yiqun Liu, Yingping Mou, and Shiping Feng}
\email{spfeng@bnu.edu.cn}

\affiliation{Department of Physics, Beijing Normal University, Beijing 100875, China}

\begin{abstract}
The recent experiments revealed a remarkable possibility for the absence of the disparity between
the phase diagrams of the electron- and hole-doped cuprate superconductors, while such an aspect
should be also reflected in the dressing of the electrons. Here the phase diagram of the
electron-doped cuprate superconductors and the related exotic features of the anisotropic
dressing of the electrons are studied based on the kinetic-energy driven superconductivity. It is
shown that although the optimized $T_{\rm c}$ in the electron-doped side is much smaller than
that in the hole-doped case, the electron- and hole-doped cuprate superconductors rather resemble
each other in the doping range of the superconducting dome, indicating an absence of the disparity
between the phase diagrams of the electron- and hole-doped cuprate superconductors. In particular,
the anisotropic dressing of the electrons due to the strong electron's coupling to a strongly
dispersive spin excitation leads to that the electron Fermi surface is truncated to form the
disconnected Fermi arcs centered around the nodal region. Concomitantly, the dip in the
peak-dip-hump structure of the quasiparticle excitation spectrum is directly associated with the
corresponding peak in the quasiparticle scattering rate, while the dispersion kink is always
accompanied by the corresponding inflection point in the total self-energy, as the dip in the
peak-dip-hump structure and dispersion kink in the hole-doped counterparts. The theory also
predicts that both the normal and anomalous self-energies exhibit the well-pronounced low-energy
peak-structures.
\end{abstract}



\maketitle

\section{Introduction}\label{Introduction-1}

The parent compound of cuprate superconductors is classified as a half-filled Mott insulator with
an antiferromagnetic (AF) long-range order (AFLRO) \cite{Bednorz86,Tokura89}, which occurs to be
due to the very strong electron correlation \cite{Anderson87}. In the hole-doped case
\cite{Bednorz86}, a small number of the doped holes destroys AFLRO quickly, and then
superconductivity appears leaving only the AF short-range order (AFSRO) correlation still intact
\cite{Fujita12,Damascelli03,Campuzano04,Fink07}. However, in the electron-doped side
\cite{Tokura89}, both the doped electrons and annealing process in a low-oxygen environment are
required to induce superconductivity, where the early experimental observations
\cite{Armitage10,Armitage01,Matsui05,Matsui05a,Matsui07,Santander-Syro11,Song12} revealed that in
a clear contrast to the hole-doped case, AFLRO survives until superconductivity appears only in a
narrow window of the highly doped electrons, around the optimal doping, and can persist into the
superconducting (SC) phase. In particular, a large pseudogap opens at around the crossing points
of the electron Fermi surface (EFS) with the AF Brillouin zone boundary due to the AFLRO
correlation \cite{Armitage10,Armitage01,Matsui05,Matsui05a,Matsui07,Santander-Syro11,Song12}.
Moreover, unlike the universal dome-like shape doping dependence of the SC transition temperature
$T_{\rm c}$ in the hole-doped case \cite{Damascelli03,Campuzano04,Fink07} with an optimal doping
at around $15\%$, the SC phase in the electron-doped side shows a large variation
\cite{Armitage10}. This disparity between the phase diagrams of the electron- and hole-doped
cuprate superconductors implies that the electron doping and hole doping may affect the electronic
structure in different manners
\cite{Damascelli03,Campuzano04,Fink07,Armitage10,Armitage01,Matsui05,Matsui05a,Matsui07,Santander-Syro11,Song12}.

Since the additional annealing process is crucial for realizing superconductivity in the
electron-doped side
\cite{Armitage10,Armitage01,Matsui05,Matsui05a,Matsui07,Santander-Syro11,Song12}, the different
annealing methods may result in distinct properties, reflecting a fact that the controversy of the
phase diagram in the electron-doped side may be attributed to the conflicting experimental results
associated with the annealing effect. Although the effect of the annealing is still not fully
understood on the microscopic level \cite{Adachi13,Adachi17}, it is possible that the intrinsic
aspects of the electron-doped cuprate superconductors are masked by the improper annealing
condition. Recently, the substantial improvements in the materials synthesis technique
\cite{Adachi17,Horio16,Song17,Lin20,Horio20b} allow one to grow the single crystals of the
electron-doped cuprate superconductors in the optimal annealing condition, where a strongly
localized state of charge carriers accompanied by an AF pseudogap at around the crossing points
in the improper annealing condition has been changed to a metallic- and SC-states with the optimal
annealing condition. In particular, the experimental observations on these new single crystals
in the proper annealing condition indicate clearly that the annealing and oxygen vacancy induce
a sufficient change in the charge carrier density \cite{Adachi17}. In this case, the doping
concentration should be considered in conjunction with the annealing and oxygen nonstoichiometry.
This actual charge carrier concentration has been be used in building the new phase diagram
\cite{Adachi17,Horio16,Song17,Lin20,Horio20b}, where the SC-state that coexists with AFSRO is
extended over a wide electron doping range with the maximum value of $T_{\rm c}$ that occurs at
around the optimal doping $\delta\sim 0.15$, while the deduced AFLRO phase boundary does not
extend into the SC dome \cite{Horio16,Song17,Lin20,Horio20b}. This new phase diagram of the
electron-doped cuprate superconductors is in a striking analogy to the corresponding phase
diagram in the hole-doped counterparts, and therefore reveals a possibility for the absence of
the disparity between the phase diagrams of the electron- and hole-doped cuprate superconductors.
With this new phase diagram, a critical question is whether or not SC mechanism in the
electron-doped side is the same as in the hole-doped case.

The new phase diagram in the electron-doped cuprate superconductors is thus closely related to
the actual electron doping concentration, while such an aspect should be reflected in the
nature of the quasiparticle excitations resulting of the dressing of the electrons due to the
electron interaction mediated by various bosonic excitations. The understanding of how strong
coupling of the electrons with various bosonic excitations affects the electronic structure is
especially important to explain the astonishing phenomena, including the question of the
pairing mechanism \cite{Vekhter03,Carbotte11,Choi18}. Very recently, the intrinsic
EFS reconstruction and its evolution with the electron doping in the new single crystals with
the proper annealing condition have been observed experimentally from the angle-resolved
photoemission spectroscopy (ARPES) measurements \cite{Horio16,Song17,Lin20,Horio20b}, where
the characteristic features can be summarized as: (a) a quasiparticle peak is observed on the
entire EFS without the signature of an AF pseudogap at around the crossing points; and (b) the
stronger quasiparticle scattering is observed in the antinodal region than in the nodal region,
leading to the formation of the disconnected Fermi arcs centered around the nodal region. In
the hole-doped case, the intrinsic features of the quasiparticle excitations, associated with
the EFS reconstruction and characterized by the distinct depression in the electron density
of states, reminiscent of the well-known peak-dip-hump (PDH) structure in the quasiparticle
excitation spectrum, the kink in the quasiparticle dispersion, and the multiple
nearly-degenerate electronic orders, have been observed experimentally by virtue of systematic
studies using the scanning tunneling microscopy \cite{Neto14,Fujita14} and resonant X-ray
scattering techniques \cite{Vishik18,Comin16,Comin14}, particularly the ARPES measurements
\cite{Dessau91,Norman97,Campuzano99,Wei08,DMou17,Kaminski01,Lanzara01,Kordyuk06,Iwasawa08,Plumb13}.
On the other hand, although a few experimental results for the dispersion kinks along the nodal
and antinodal directions, associated with the new phase diagram and the intrinsic EFS
reconstruction, has been observed very recently in the electron-doped side with the proper
annealing condition \cite{Horio20b}, the experimental data of the PDH structure are still
lacking to date. Furthermore, to the best of our knowledge, these intrinsic properties of the
quasiparticle excitations have also not been discussed starting from a SC theory so far. In
this case, the crucial issue is to understand even from a theoretical analysis which intrinsic
aspects of the quasiparticle excitations are universal to both the electron- and hole-doped
cuprate superconductors, and how they might depend on the specifics of the participating
electron- or hole-like states.

In our previous works \cite{Feng15a,Gao18,Gao18a,Gao19,Liu20}, the doping dependence of
$T_{\rm c}$ and the related dressing of the electrons in the hole-doped cuprate superconductors
have been investigated within the framework of the kinetic-energy driven superconductivity, where
we have shown that $T_{\rm c}$ takes a dome-like shape with the underdoped and overdoped regimes
on each side of the optimal doping $\delta\sim 0.15$, where $T_{\rm c}$ reaches its maximum, and
then the main aspects of the quasiparticle excitations observed from the experiments,
including the EFS reconstruction \cite{Norman98,Kanigel07,Nakayama09,Kondo13,Chatterjee06}, the
nature of the charge-order correlation \cite{Neto14,Fujita14,Vishik18,Comin16,Comin14}, the
striking PDH structure in the quasiparticle excitation spectrum
\cite{Dessau91,Norman97,Campuzano99,Wei08,DMou17}, the dispersion kink
\cite{Kaminski01,Lanzara01,Kordyuk06,Iwasawa08,Plumb13}, and the remarkable ARPES autocorrelation
and its connection with the quasiparticle scattering interference \cite{Chatterjee06,He14}, are
qualitatively reproduced. In particular, we have also shown that all these exotic features are a
natural result of the strong electron correlation characterized by the strong electron interaction
mediated by a strongly dispersive spin excitation. However, a comprehensive discussion of the
electron-doped counterparts has not been given. We believe that if the coupling of the electrons
with a strongly dispersive spin excitation is of the same nature both in the electron- and
hole-doped cuprate superconductors, it should reveal itself in the nature of the quasiparticle
excitation of the electron-doped cuprate superconductors as it does in the hole-doped counterparts.
In this paper, we try to study the phase diagram and the related intrinsic properties of the
quasiparticle excitations in the electron-doped cuprate superconductors. We show explicitly that
the maximal $T_{\rm c}$ occurs around the optimal doping $\delta\sim 0.15$, and then decreases in
both the underdoped and overdoped regimes. In particular, although the optimized $T_{\rm c}$ in the
electron-doped side is much smaller than that in the hole-doped case, the electron- and hole-doped
cuprate superconductors rather resemble each other in the doping range of the SC dome, indicating
an absence of the disparity between the phase diagrams of the electron- and hole-doped cuprate
superconductors. Moreover, the characteristic features of the intrinsic EFS reconstruction, the
PDH structure, and the dispersion kink in the electron-doped cuprate superconductors are a clear
analogy to those obtained in the hole-doped counterparts. Our present results therefore also show
that the essential physics is the same for both the electron- and hole-doped cuprate
superconductors.

This paper is organized as follows. First, we present the basic formalism in Sec. \ref{Formalism},
where we express explicitly the single-particle diagonal and off-diagonal propagators (hence the
single-particle spectral function) of the electron-doped cuprate superconductors based on the
kinetic-energy driven superconductivity. We then discuss the doping dependence of $T_{\rm c}$ in
Sec. \ref{phase-diagram}, where a comparison of the phase diagrams between the electron- and
hole-doped cuprate superconductors is made. In Sec. \ref{electronic-state}, we discuss the
intrinsic aspects of the quasiparticle excitations, and show that a sharp quasiparticle
peak emerges on the entire EFS without an AF pseudogap at around the crossing points. Finally, we
give a summary and discussions in Sec. \ref{conclusions}.

\section{Formalism}\label{Formalism}

Both the electron- and hole-doped cuprate superconductors have a layered crystal structure
consisting of the two-dimensional CuO$_{2}$ planes separated by insulating layers
\cite{Bednorz86,Tokura89}, and it is then believed that the exotic features in these systems are
closely related to the doped CuO$_{2}$ planes \cite{Damascelli03,Campuzano04,Fink07,Armitage10}.
In this case, as originally emphasized by Anderson \cite{Anderson87}, the essential physics of the
doped CuO$_{2}$ plane is properly captured by the $t$-$J$ model on a square lattice
\cite{Lee06,Edegger07} ,
\begin{eqnarray}\label{tjham}
H&=&-t\sum_{l\hat{\eta}\sigma}C^{\dagger}_{l\sigma}C_{l+\hat{\eta}\sigma}
+t'\sum_{l\hat{\tau}\sigma}C^{\dagger}_{l\sigma}C_{l+\hat{\tau}\sigma} \nonumber\\
&+&\mu\sum_{l\sigma}C^{\dagger}_{l\sigma}C_{l\sigma}
+J\sum_{l\hat{\eta}}{\bf S}_{l}\cdot {\bf S}_{l+\hat{\eta}},~~
\end{eqnarray}
where the summation is over all sites $l$, and for each $l$, over its nearest-neighbors (NN)
$\hat{\eta}$ or the next NN $\hat{\tau}$, the hopping integrals $t>0$ and $t'>0$ for the hole-doped
case, while $t<0$ and $t'<0$ in the electron-doped side. In particular, the NN hopping $t$ in the
$t$-$J$ model (\ref{tjham}) has a electron-hole symmetry because the sign of $t$ can be absorbed by
the change of the sign of the orbital on one sublattice. However, the electron-hole asymmetry can be
properly accounted by the next NN hopping $t'$ \cite{Hybertsen90,Gooding94,Kim98}.
$C^{\dagger}_{l\sigma}$ and $C_{l\sigma}$ are the electron creation and annihilation operators,
respectively, with spin $\sigma$ on site $l$, ${\bf S}_{l}$ is a localized spin operator, and $\mu$
is the chemical potential. The high complexity in the $t$-$J$ model (\ref{tjham}) comes mainly from
the electron local constraint, i.e., this $t$-$J$ model (\ref{tjham}) is supplemented by a local
constraint of no double electron occupancy in the hole-doped case:
$\sum_{\sigma}C^{\dagger}_{l\sigma}C_{l\sigma}\leq 1$, while it acts on the space with no zero
electron occupied sites in the electron-doped side:
$\sum_{\sigma}C^{\dagger}_{l\sigma}C_{l\sigma}\geq 1$. However, for the electron doping, we can
work in the hole representation via a particle-hole transformation
$C_{l\sigma}\rightarrow f^{\dagger}_{l-\sigma}$, with $f^{\dagger}_{l\sigma}$ ($f_{l\sigma}$) that
is the hole creation (annihilation) operator, and then the local constraint of no zero electron
occupancy in the electron representation $\sum_{\sigma}C^{\dagger}_{l\sigma}C_{l\sigma}\geq 1$ is
replaced by the local constraint of no double hole occupancy in the hole representation
$\sum_{\sigma}f^{\dagger}_{l\sigma}f_{l\sigma}\leq 1$. In this case, the $t$-$J$ model (\ref{tjham})
in both the electron doping and hole doping is always subject to an important on-site local
constraint to avoid the double occupancy, and the difference between the electron doping and hole
doping is expressed as the sign difference of the hopping integrals as mentioned above. This local
constraint of no double occupancy can be dealt properly by the fermion-spin theory
\cite{Feng9404,Feng15} based on the charge-spin separation, and in particular, the constrained hole
operators $f_{l\uparrow}$ and $f_{l\downarrow}$ are decoupled as,
\begin{eqnarray}\label{CSS}
f_{l\uparrow}=a^{\dagger}_{l\uparrow}S^{-}_{l}, ~~~~
f_{l\downarrow}=a^{\dagger}_{l\downarrow}S^{+}_{l},
\end{eqnarray}
where the spinful fermion operator $a_{l\sigma}=e^{-i\Phi_{l\sigma}}a_{l}$ describes the charge
degree of freedom of the constrained hole together with some effects of spin configuration
rearrangements due to the presence of the doped charge carrier itself, while the localized spin
operator $S_{l}$ describes the spin degree of freedom of the constrained hole, and then the local
constraint without double hole occupancy is satisfied in analytical calculations. Based on the
$t$-$J$ model in the fermion-spin representation, the kinetic-energy driven SC mechanism has been
established \cite{Feng15,Feng0306,Feng12}, where in the doped regime without an AFLRO, the coupling
of the charge carriers and spin excitations directly from the kinetic energy provides the attractive
interaction between the charge carriers in the particle-particle channel that pairs charge carriers
together to form d-wave charge-carrier pairing state, then the electron pairs with the d-wave
symmetry originated from the d-wave charge-carrier pairing state are due to the charge-spin
recombination \cite{Feng15a}, and their condensation reveals the SC ground-state. The typical
features of the kinetic energy driven SC mechanism can be summarized as: (a) the mechanism is
purely electronic without phonons; (b) the mechanism indicates that the strong electron correlation
favors superconductivity, since the main ingredient is identified into an electron pairing mechanism
not involving the phonon, the external degree of freedom, but the internal spin degree of freedom of
electron; (c) the electron pairing state is controlled by both the electron pair gap and
single-particle coherence, which leads to that the maximal $T_{\rm c}$ occurs around the optimal
doping, and then decreases in both the underdoped and the overdoped regimes; (d) superconductivity
coexists with the AFSRO correlation. Within the framework of this kinetic-energy driven
superconductivity, the ARPES autocorrelation and the
line-shape of the quasiparticle excitation spectrum in electron-doped cuprate superconductors have been
studied recently \cite{Tan20}. Following these previous discussions \cite{Tan20}, the single-particle
diagonal and off-diagonal propagators $G({\bf k},\omega)$ and $\Im^{\dagger}({\bf k},\omega)$ of the
electron-doped cuprate superconductors can be obtained explicitly as,
\begin{subequations}\label{EGF}
\begin{eqnarray}
G({\bf k},\omega)&=&{1\over\omega-\varepsilon_{\bf k}-\Sigma_{\rm tot}({\bf k},\omega)},
\label{DEGF}\\
\Im^{\dagger}({\bf k},\omega)&=&{L_{\bf k}(\omega)\over \omega- \varepsilon_{\bf k}
-\Sigma_{\rm tot}({\bf k},\omega)},\label{ODEGF}
\end{eqnarray}
\end{subequations}
where $\varepsilon_{\bf k}=4t\gamma_{\bf k}-4t'\gamma_{\bf k}'-\mu$ is the single-electron band energy,
with $\gamma_{\bf k}=({\rm cos}k_{x}+{\rm cos} k_{y})/2$ and
$\gamma_{\bf k}'={\rm cos}k_{x}{\rm cos}k_{y}$, $L_{\bf k}(\omega)=-\Sigma_{\rm pp}({\bf k},\omega)
/[\omega+\varepsilon_{\bf k}+\Sigma_{\rm ph}({\bf k},-\omega)]$, while the total self-energy
$\Sigma_{\rm tot}({\bf k},\omega)$ is a combination of the normal self-energy
$\Sigma_{\rm ph}({\bf k},\omega)$ in the particle-hole channel and the anomalous self-energy
$\Sigma_{\rm pp} ({\bf k},\omega)$ in the particle-particle channel, and is given explicitly by,
\begin{eqnarray}
\Sigma_{\rm tot}({\bf k},\omega)=\Sigma_{\rm ph}({\bf k},\omega)
+{|\Sigma_{\rm pp}({\bf k},\omega)|^{2}\over\omega+\varepsilon_{\bf k}
+\Sigma_{\rm ph}({\bf k},-\omega)}. \label{TOT-SE}
\end{eqnarray}
In the kinetic-energy driven SC mechanism, both the normal and anomalous self-energies
$\Sigma_{\rm ph}({\bf k},\omega)$ and $\Sigma_{\rm pp}({\bf k},\omega)$ quantify the interaction
between electrons mediated by a strongly dispersive spin excitation, and have been given explicitly
in Ref. \onlinecite{Tan20}, where the sharp peak visible for temperature $T\rightarrow 0$ in the
normal (anomalous) self-energy is actually a $\delta$-functions, broadened by a small damping used
in the numerical calculation at a finite lattice. The calculation in this paper for the normal
(anomalous) self-energy is performed numerically on $160\times 160$ lattice in momentum space, with
the infinitesimal $i0_{+}\rightarrow i\Gamma$ replaced by a small damping $\Gamma=0.1J$.

The result in Eq. (\ref{EGF}) also shows that the basic Eliashberg formalism \cite{Eliashberg60}
with the d-wave type SC gap is still valid, although the pairing mechanism is driven by the kinetic
energy by the exchange of spin excitations. The single-particle spectral function
$A({\bf k},\omega)$ is related directly to the imaginary part of the single-particle diagonal
propagator in Eq. (\ref{DEGF}) as \cite{Damascelli03,Campuzano04,Fink07},
\begin{eqnarray}\label{ESF}
A({\bf k},\omega)={-2{\rm Im}\Sigma_{\rm tot}({\bf k},\omega)\over [\omega-\varepsilon_{\bf k}
-{\rm Re}\Sigma_{\rm tot}({\bf k},\omega)]^{2}+[{\rm Im} \Sigma_{\rm tot}({\bf k},\omega)]^{2}},~
\end{eqnarray}
and then the quasiparticle excitation spectrum measured by ARPES experiments is proportional to
this single-particle spectral function (\ref{ESF}), where ${\rm Re}\Sigma_{\rm tot}({\bf k},\omega)$
and ${\rm Im}\Sigma_{\rm tot}({\bf k},\omega)$ are the real and imaginary parts of the total
self-energy $\Sigma_{\rm tot}({\bf k},\omega)$, respectively. In particular, the real part of the
total self-energy offsets the single-electron band energy $\varepsilon_{\bf k}$, while the imaginary
part of the total self-energy is identified as the quasiparticle scattering rate,
\begin{eqnarray}\label{rate}
\Gamma({\bf k},\omega)={\rm Im}\Sigma_{\rm tot}({\bf k},\omega)
\end{eqnarray}
which broadens the spectral peak in the ARPES spectrum, and therefore governs the lifetime of the
quasiparticle \cite{Damascelli03,Campuzano04,Fink07}. This is why one can obtain the information about the
total self-energy from ARPES experiments by analyzing the energy and momentum distribution data.

\section{Phase diagram} \label{phase-diagram}

\begin{figure}[h!]
\centering
\includegraphics[scale=0.90]{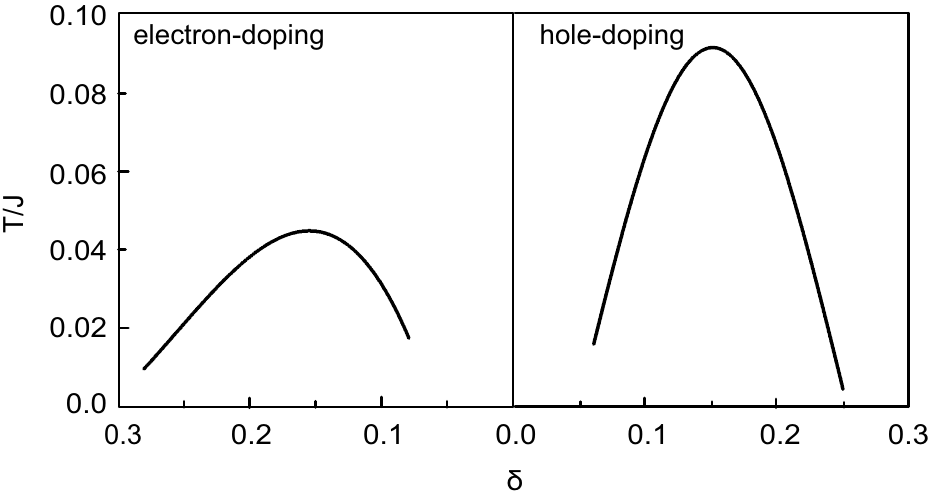}
\caption{Left panel: the doping dependence of the superconducting transition-temperature in
the electron-doped side for $t/J=-2.5$ and $t'/t=0.3$. Right panel: the corresponding result
in the hole-doped case for $t/J=2.5$ and $t'/t=0.3$ taken from Ref. \onlinecite{Feng15a}.
\label{Tc-doping}}
\end{figure}

The understanding of the phase diagram of cuprate superconductors has been one of the central issues
since its original discovery in 1986 \cite{Bednorz86,Tokura89}, through to the present day. Within
the framework of the kinetic-energy driven superconductivity \cite{Feng15,Feng0306,Feng12}, the
evolution of $T_{\rm c}$ with the hole doping in the hole-doped cuprate superconductors has been
obtained in our early studies \cite{Feng15a,Feng12} in terms of the self-consistent calculation at
the condition of the SC gap $\bar{\Delta}=0$, where $T_{\rm c}$ has a dome-like shape doping
dependence with the maximum $T_{\rm c}$ that occurs at around the optimal hole doping
$\delta\sim 0.15$, in good agreement with the experimental results observed on the hole-doped cuprate
superconductors \cite{Tallon95}. Following these early studies for the hole-doped case, we have also
performed a self-consistent calculation for the doping dependence of $T_{\rm c}$ in the electron-doped
side, and the result of $T_{\rm c}$ as a function of the electron doping for parameters $t/J=-2.5$ and
$t'/t=0.3$ is plotted in the left panel of Fig. \ref{Tc-doping}. In order to compare the present result
of the electron-doped cuprate superconductors with that of the hole-doped counterparts, the
corresponding result \cite{Feng15a} of the doping dependence of $T_{\rm c}$ in the hole-doped case for
parameters $t/J=2.5$ and $t'/t=0.3$ is also shown in the right panel of Fig. \ref{Tc-doping}. One can
immediately see from the results in Fig. \ref{Tc-doping} that the doping range of the SC dome in the
electron-doped side is very similar to that in the hole-doped case, where with the increase of doping,
$T_{\rm c}$ is raised gradually in the underdoped regime, and reaches the maximum around the optimal
doping $\delta\sim 0.15$, however, the optimized $T_{\rm c}$ in the electron-doped side is much lower
than that in the hole-doped case. With the further increase of doping, $T_{\rm c}$ then turns into a
monotonically decrease in the overdoped regime, and finally, superconductivity disappears in the
heavily overdoped regime. Apart from this obvious similarity of the doping ranges of the SC domes in
the electron- and hole-doped cuprate superconductors, the other typical features in
Fig. \ref{Tc-doping} can be also summarized as: (a) there is no the disparity between the phase diagrams
of the electron- and hole-doped cuprate superconductors; (b) superconductivity coexists with the AFSRO
correlation; (c) as an evidence of the electron-hole asymmetry, $T_{\rm c}$ in the hole-doped case is
much higher than that in the electron-doped side. This electron doping dependence of $T_{\rm c}$ in the
left panel of Fig. \ref{Tc-doping} and the related typical features are well consistent with the true
phase diagram observed recently on Pr$_{1-x}$LaCe$_{x}$CuO$_{4-\delta}$ under the proper annealing
condition \cite{Song17,Lin20} and the corresponding $\mu$SR experimental results \cite{Adachi16}. The
good agreement between the present theoretical result of the phase diagram in Fig. \ref{Tc-doping} and
experimental observations also show the existence of a common SC mechanism for both the electron- and
hole-doped cuprate superconductors.

\begin{figure}[h!]
\centering
\includegraphics[scale=0.75]{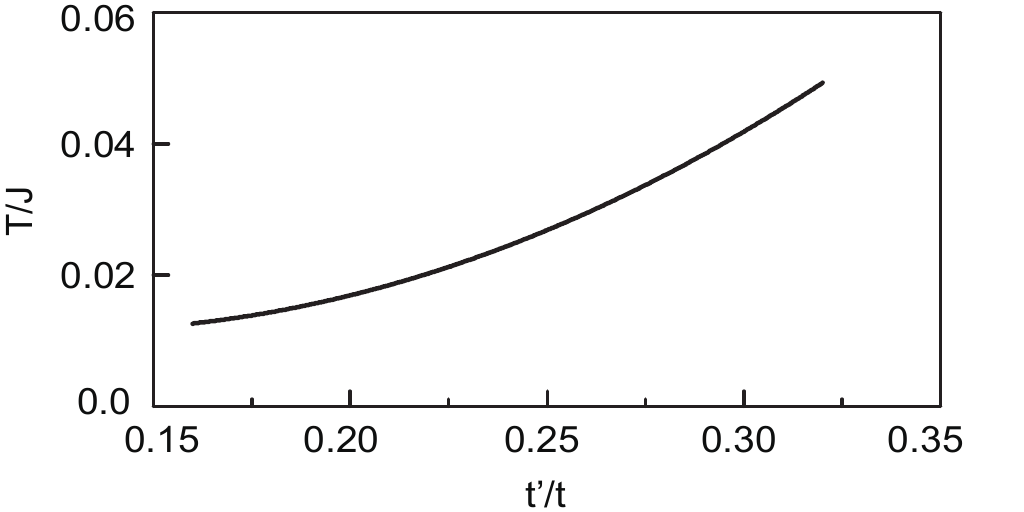}
\caption{The maximal superconducting transition-temperature as a function of $t'$ at the electron doping
$\delta=0.15$ for $t/J=-2.5$. \label{Tc-t1}}
\end{figure}

However, it should be noted that both in the electron- and hole-doped cuprate superconductors, the
values of $t$, $t'$, and $J$ are believed to vary somewhat from compound to compound
\cite{Hybertsen90,Gooding94,Kim98,Park07,Nekrasov09,Ikeda09,Ikeda09a,Harter12,Harter15,Tanaka04,Shih04,Pavarini01},
and then some differences among the different families such as the significant variation in the
maximal $T_{\rm c}$ at the optimal doping have been observed experimentally
\cite{Damascelli03,Campuzano04,Fink07,Armitage10}. In particular, it has been shown experimentally and
theoretically that the maximal $T_{\rm c}$ (then the shape of the SC dome) for different families of
the hole-doped cuprate superconductors \cite{Tanaka04,Shih04,Pavarini01} is strongly correlated with
$t'$. In this case, we have made a series of calculations for the maximal $T_{\rm c}$ at the optimal
doping with different values of $t'$ in the electron-doped cuprate superconductors, and the results
show that the maximal $T_{\rm c}$ is also correlated with $t'$. To see this correlation more clearly,
we plot the maximal $T_{\rm c}$ as a function of $t'$ at the electron doping $\delta=0.15$ for
$t/J=-2.5$ in Fig. \ref{Tc-t1}, where in a reasonably estimative range
\cite{Hybertsen90,Gooding94,Kim98,Park07,Nekrasov09,Ikeda09,Ikeda09a,Harter12,Harter15} of the
parameter $t'$, the maximal $T_{\rm c}$ is enhanced by $t'$, i.e., the maximal $T_{\rm c}$ increases
smoothly with the increase of $t'$. This anticipated result is also well consistent with the
corresponding experimental data \cite{Tanaka04} of the hole-doped counterparts, and the numerical
simulations \cite{Shih04,Pavarini01} in the same range of the parameter $t'$, where the enhancement
of $T_{\rm c}$ by $t'$ has been observed experimentally and confirmed by the numerical simulations.

\section{Exotic features of dressing of electrons} \label{electronic-state}

The dressing of the electrons due to the strong coupling of the electrons with various bosonic
excitations leads to a rich variety phenomena. However, with the new phase diagram in
Fig. \ref{Tc-doping}, the immediate problem becomes to study which these astonishing phenomena
that are universal to both the electron- and hole-doped cuprate superconductors, and how these
astonishing phenomena depend on the specifics of the participating electron- or hole-like states.

\subsection{Intrinsic electron Fermi surface reconstruction}\label{Fermi-arcs}

\begin{figure}[h!]
\centering
\includegraphics[scale=0.83]{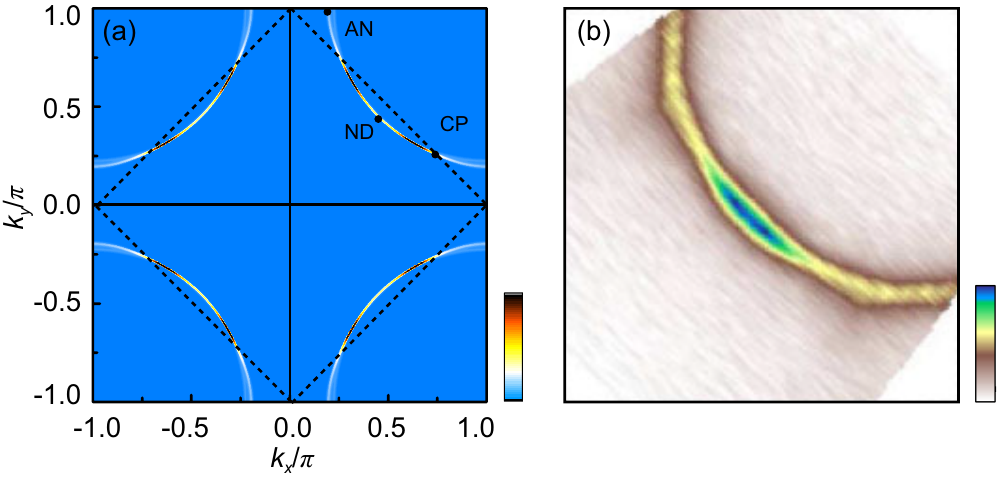}
\caption{(Color online) (a) The electron Fermi surface map at the electron doping $\delta=0.15$
with $T=0.002J$ for $t/J=-2.5$ and $t'/t=0.3$, where the dotted-line represents the
antiferromagnetic Brillouin zone boundary, while AN, CP, and ND denote the antinode, crossing
point, and node, respectively. (b) The experimental result of the electron Fermi surface for
Pr$_{1-x}$LaCe$_{x}$CuO$_{4-\delta}$ under the proper annealing at around the electron doping
$\delta=0.15$ taken from Ref. \onlinecite{Song17}. \label{EFS-map}}
\end{figure}

In ARPES experiments \cite{Damascelli03,Campuzano04,Fink07}, the weight of ARPES spectrum at
zero energy is used to map out EFS. In other words, EFS is determined by the single-particle
spectral function $A({\bf k},\omega=0)$ in Eq. (\ref{ESF}) at zero energy to map out the locus
of the maximum in the spectral weight of $A({\bf k},\omega=0)$. For the understanding of the
essential physics of an interacting system, the study of the nature of EFS is a starting point
\cite{Mahan81}. This follows from a fact that the EFS topology is a fundamental ground-state
property, and influences directly the low-energy transport properties. In particular, the
shape of EFS in cuprate superconductors has been central to addressing the strong electron
correlation effect and multiple nearly-degenerate electronic orders
\cite{Neto14,Fujita14,Vishik18,Comin16,Comin14,Neto15,Neto16}. In this case, we first
characterize the EFS topology. In the case without considering the electron interaction, a
large EFS is characterized as a closed contour of the gapless quasiparticle excitations in
momentum space, where the peaks of the quasiparticle excitation spectrum with the same weight
distribute uniformly along with the EFS contour, and then the quasiparticle lifetime is
infinitely long on this EFS contour. However, in the presence of the strong interaction of
the electrons with a strongly dispersive spin excitation, which is manifested itself by the
energy and momentum dependence of the total self-energy $\Sigma_{\rm tot}({\bf k},\omega)$,
the closed EFS contour is broken up into the disconnected Fermi arcs. To see this EFS
reconstruction more clearly, we plot the underlying EFS map in the $[k_{x},k_{y}]$ plane at
the electron doping $\delta=0.15$ with $T=0.002J$ for $t/J=-2.5$ and
$t'/t=0.3$ in Fig. \ref{EFS-map}a. For comparison, the experimental result \cite{Song17}
obtained from the ARPES measurement on Pr$_{1-x}$LaCe$_{x}$CuO$_{4-\delta}$ under the proper
annealing at around the electron doping $\delta=0.15$ is also shown in Fig. \ref{EFS-map}b.
In a qualitative analogy to the hole-doped counterparts \cite{Feng15a,Gao18}, EFS has been
reconstructed due to the strong redistribution of the spectral weight, where there are two
typical regions: (a) the antinodal region, where the result shows a heavy suppression
of the spectral weight, leading to that EFS at around the antinodal region is invisible.
Moreover, the AF pseudogap at around the crossing points is totally absent; (b) the nodal
region, where the result indicates a modest reduction of the spectral weight, and then EFS is
clearly visible as the reminiscence of the EFS contour in the case without considering the
electron interaction to form the Fermi arcs centered around the nodal region. However, the
dressing from the quasiparticle scattering further moves the spectral weight from the Fermi
arcs to the tips of the Fermi arcs, which leads to that although a quasiparticle peak
emerges in the entire EFS, the spectral weight exhibits a largest value at around the tips
of the Fermi arcs. This EFS reconstruction is also qualitatively consistent with the recent
experimental observations on the electron-doped cuprate superconductors
\cite{Horio16,Song17,Lin20,Horio20b}, where upon the proper annealing, (a) the weight of the
ARPES spectrum around the antinodal region is suppressed, and then EFS is truncated to form
the Fermi arcs located around the nodal region; (b) the AF pseudogap is closed on the whole
EFS, and then the quasiparticle peak is observed on the whole EFS. Moreover, as in the
hole-doped case \cite{Gao19,Chatterjee06,He14}, these tips of the Fermi arcs linked up by the
scattering wave vectors ${\bf q}_{i}$ in the electron-doped side also can construct an
{\it octet} scattering model \cite{Tan20}, and then different ordered electronic states
determined by the quasiparticle scattering processes with the corresponding scattering wave
vectors ${\bf q}_{i}$ therefore are driven by this intrinsic EFS instability.
This is also why a exotic feature in the electron-doped cuprate superconductor is the
coexistence and competition between different ordered electronic states and superconductivity
\cite{Armitage10,Neto15,Neto16}. In particular, it has been shown that the charge-order wave
vector ${\bf q}_{1}={\bf Q}_{\rm co}$ connecting the straight tips of the Fermi arcs smoothly
{\it increases} with the increase of the electron doping \cite{Mou17}, which is consistent with
the experimental results \cite{Neto16}. However, this is unlike the case in the hole-doped
counterparts \cite{Gao18,Comin16,Comin14}, where the charge-order wave vector smoothly
{\it decreases} with the increase of the hole doping, which is another evidence of the
electron-hole asymmetry. Furthermore, we have also studied the doping dependence of the EFS
reconstruction, and the results show that the EFS reconstruction can persist into the overdoped
regime, also in agreement with the experimental observations \cite{Horio16,Song17,Lin20,Horio20b}.

\begin{figure}[h!]
\centering
\includegraphics[scale=1.0]{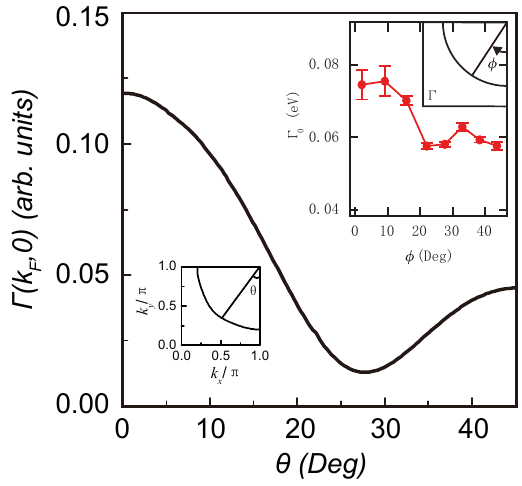}
\caption{(Color online) The angular dependence of the quasiparticle scattering rate along
${\bf k}_{\rm F}$ from the antinode to the node at $\delta=0.15$ with $T=0.002J$ for
$t/J=-2.5$ and $t'/t=0.3$. Inset: the corresponding experimental data of
Pr$_{1.3-x}$La$_{0.7}$Ce$_{x}$CuO$_{4}$ under the proper annealing condition taken from
Ref. \onlinecite{Horio16}. \label{scattering-rate}}
\end{figure}

The physical origin of the EFS reconstruction in the electron-doped side is the same as that in
the hole-doped case \cite{Gao18}, and can be also attributed to the highly anisotropic momentum
dependence of the quasiparticle scattering rate (\ref{rate}). The EFS contour in momentum space
is determined directly by the poles of the single-particle diagonal propagator (\ref{DEGF}) at
zero energy: $\varepsilon_{\bf k}+{\rm Re}\Sigma_{\rm tot}({\bf k},0)=0$, and then the spectral
weight of the single-particle spectral function $A({\bf k},0)$ in Eq. (\ref{ESF}) at EFS is
governed by the inverse of the quasiparticle scattering rate $1/\Gamma({\bf k},0)$. To reveal
this highly anisotropic $\Gamma({\bf k},0)$ in momentum space clearly, we plot the angular
dependence of $\Gamma({\bf k}_{\rm F},0)$ along EFS from the antinode to the node at $\delta=0.15$
with $T=0.002J$ for $t/J=-2.5$ and $t'/t=0.3$ in Fig. \ref{scattering-rate} in comparison with the
corresponding ARPES result \cite{Horio16} observed on Pr$_{1.3-x}$La$_{0.7}$Ce$_{x}$CuO$_{4}$
under the proper annealing condition (inset). Apparently, the main feature of the quasiparticle
scattering rate along EFS observed from the ARPES experiment \cite{Horio16} is qualitatively
reproduced, where the quasiparticle scattering rate increases as the momentum approaches the
antinode, and then the strongest scattering emerges at the antinode, leading to a heavy suppression
of the spectral weight at around the antinode. Moreover, the weak quasiparticle scattering occurs
around the nodal region, and then the formation of the Fermi arcs is due to a modest
reduction of the spectral weight around the nodal region. This strong redistribution of the
spectral weight on EFS therefore induces the EFS reconstruction or instability. It is very
interesting to note that the similar angular dependence of the quasiparticle scattering rate has
been also observed experimentally in the hole-doped case \cite{Vishik09}, indicating a common
quasiparticle scattering mechanism both in the hole- and electron-doped cuprate superconductors.

However, as seen by comparison with the experimental data \cite{Horio16}, there is also a
substantial difference between theory and experiment, namely, the weakest quasiparticle scattering
occurs at the crossing points in experiment \cite{Horio16}, while the calculation in the present
parameters $t/J=-2.5$ and $t'/t=0.3$ anticipates it at the tips of the Fermi arcs. In the
electron-doped cuprate superconductors, (a) as we have mentioned above in Section
\ref{phase-diagram}, the values of $t$, $t'$, and $J$ are different between different families
\cite{Park07,Nekrasov09,Ikeda09,Ikeda09a,Harter12,Harter15}; (b) however, the positions of the tips
of the Fermi arcs are strongly depend on the curvature of EFS, while this curvature
\cite{Park07,Nekrasov09,Ikeda09,Ikeda09a,Harter12,Harter15} is dominated by the next NN hopping
$t'$. In other words, the difference in $t'$ among the electron-doped cuprate superconductors
\cite{Park07,Nekrasov09,Ikeda09,Ikeda09a,Harter12,Harter15} leads to the difference of the positions
of the tips of the Fermi arcs. In this case, we have also made a series of calculations for the
different value of $t'$, and found that the positions of the tips of the Fermi arcs can locate
exactly on the corresponding positions of the crossing points by the proper choice or adjustment of
the parameter $t'$.

\subsection{Peak-dip-hump structure}

\begin{figure}[h!]
\centering
\includegraphics[scale=0.85]{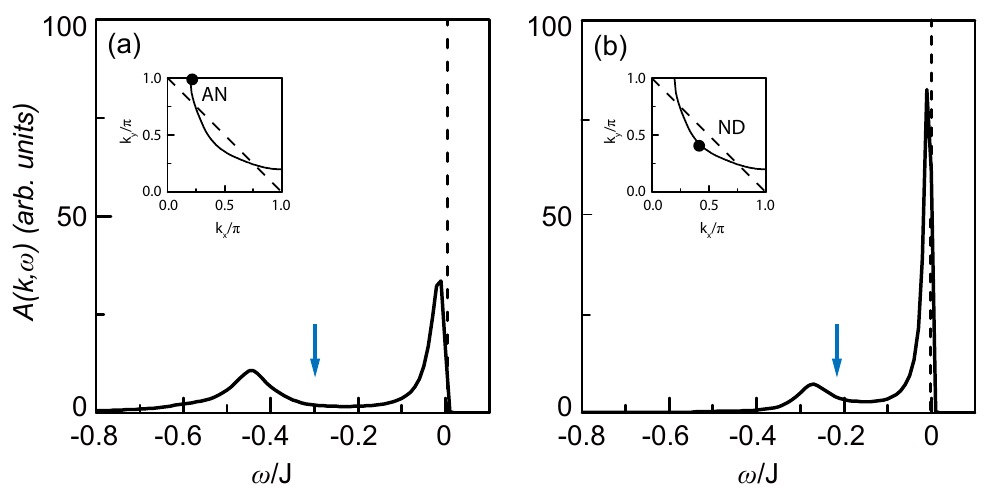}
\caption{(Color online) The quasiparticle excitation spectrum as a function of energy at (a)
the antinode and (b) the node in $\delta=0.15$ with $T=0.002J$ for $t/J=-2.5$ and $t'/t=0.3$,
where the blue arrow indicates the position of the dip, while AN and ND in the insets denote
the antinode and node, respectively. \label{ED-PDH}}
\end{figure}

The dressing of the electrons affects the quasiparticle excitation spectrum and the momentum and
energy dependence of the quasiparticle scattering rate, which can be obtained respectively from
the energy distribution curves and the widths of the corresponding peaks in ARPES
experiments \cite{Damascelli03,Campuzano04,Fink07}. In the hole-doped cuprate superconductors, a
hallmark of the spectral line-shape of the ARPES spectrum is the well-known PDH structure
\cite{Dessau91,Norman97,Campuzano99,Wei08,DMou17}, which is closely associated with the EFS
reconstruction, and now has been identified experimentally along the entire EFS. Since the strong
coupling of the electrons with bosonic excitations has been observed experimentally in whole
families of the hole-doped cuprate superconductors within a wide hole doping range, it is thus
believed that the strong coupling of the electrons with bosonic excitations gives rise to the PDH
structure \cite{Dessau91,Norman97,Campuzano99,Wei08,DMou17}. In particular, we \cite{Gao18,Gao18a}
have shown based on the kinetic-energy driven SC mechanism that this strong coupling of the
electrons with bosonic excitations can be identified as the strong electron's coupling to a
strongly dispersive spin excitation. In this subsection, we discuss the spectral line-shape in the
quasiparticle excitation spectrum of the electron-doped cuprate superconductors. We have performed
a calculation for the single-particle spectral function (\ref{ESF}), and the results of
$A({\bf k},\omega)$ as a function of energy at (a) the antinode and (b) node for $\delta=0.15$
with $T=0.002J$ for $t/J=-2.5$ and $t'/t=0.3$ are plotted in Fig. \ref{ED-PDH}, where we find that
the main features of the PDH structure in the electron-doped side are in a striking similarity to
those obtained in the hole-doped case \cite{Dessau91,Norman97,Campuzano99,Wei08,DMou17,Gao18,Gao18a}.
In particular, the position of the dip in the PDH structure is shifted to the lower energy when one
moves the momentum from the antinode to the node, while the coupling strength of the electrons with
a strongly dispersive spin excitation appears to be weaker at around the nodal region than at around
the antinodal region. Moreover, the results in Fig. \ref{ED-PDH} also show that the PDH structure is
an intrinsic feature of the quasiparticle excitation spectrum in the electron-doped cuprate
superconductors, and is found to be present all around EFS. Although the experimental data of the
PDH structure in the quasiparticle excitation spectrum of the electron-doped cuprate superconductors
under the proper annealing condition are still lacking to date, the similar PDH structure has been
observed experimentally on the electron-doped cuprate superconductors under the improper annealing
condition \cite{Armitage10,Armitage01,Matsui05,Matsui05a,Matsui07}, while our present results in
Fig. \ref{ED-PDH} are also qualitatively consistent with these experimental observations
\cite{Armitage10,Armitage01,Matsui05,Matsui05a,Matsui07}.

\begin{figure}[h!]
\centering
\includegraphics[scale=0.85]{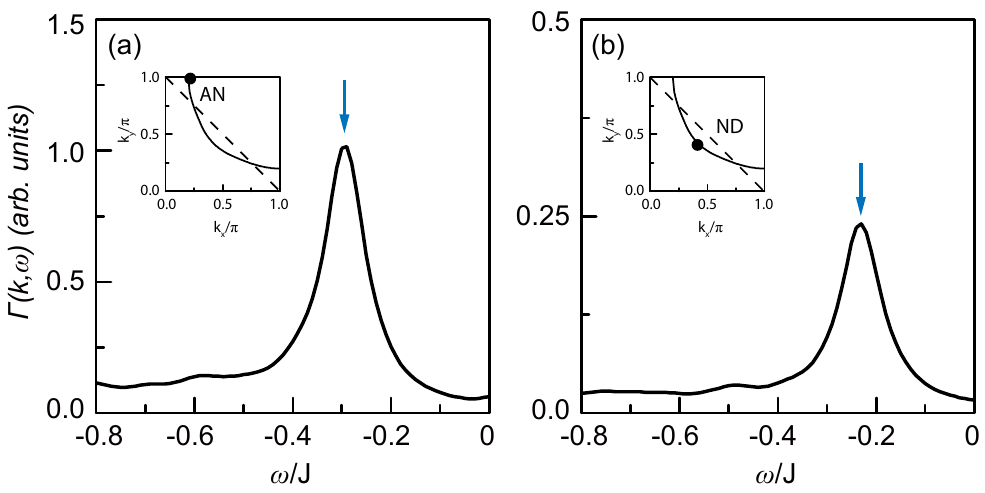}
\caption{(Color online) The quasiparticle scattering rate at (a) the antinode and (b) the node as a
function of energy in $\delta=0.15$ with $T=0.002J$ for $t/J=-2.5$ and $t'/t=0.3$, where the blue
arrow indicates the position of the peak, and AN and ND in the insets denote the antinode and node,
respectively. \label{PDH-scattering-rate}}
\end{figure}

The mechanism of the PDH structure in the quasiparticle excitation spectrum of the
electron-doped cuprate superconductors is also the same as in the hole-doped counterparts
\cite{DMou17,Gao18,Gao18a}, and is directly associated with the emergence of the corresponding
sharp peak in the quasiparticle scattering rate. Following the single-particle spectral function
in Eq. (\ref{ESF}), the position of the peak in the momentum distribution curve, plotted as a
function of energy in Fig. \ref{ED-PDH}, is determined self-consistently by the quasiparticle
dispersion,
\begin{eqnarray}\label{band}
\bar{E}_{\bf k}=\varepsilon_{\bf k}+{\rm Re}\Sigma_{\rm tot}({\bf k},\bar{E}_{\bf k}),
\end{eqnarray}
while the weight of the peak (then the lifetime of the quasiparticle excitation) is dominated by
the inverse of the quasiparticle scattering rate $\Gamma({\bf k},\omega)$. In this case, the
appearance of the dip in the PDH structure of the quasiparticle excitation spectrum along EFS is
closely related to the emergence of the corresponding sharp peak in $\Gamma({\bf k},\omega)$.
To see this point more clearly, we plot $\Gamma({\bf k},\omega)$ as a function of energy at (a)
the antinode and (b) the node for $\delta=0.15$ with $T=0.002J$ for $t/J=-2.5$ and $t'/t=0.3$ in
Fig. \ref{PDH-scattering-rate}, where the sharp peaks emerge at the antinode and node,
respectively. More specifically, we find that the position of the peak in $\Gamma({\bf k},\omega)$
at the antinode (node) in Fig. \ref{PDH-scattering-rate} is exactly corresponding to the position
of the dip in the PDH structure in the quasiparticle excitation spectrum at the antinode (node)
shown in Fig. \ref{ED-PDH}, and then the spectral weight at around the dip is suppressed heavily
by the corresponding sharp peak in $\Gamma({\bf k},\omega)$, which leads to form the PDH structure
in the quasiparticle excitation spectrum. In other words, the striking PDH structure in the
quasiparticle excitation spectrum shown in Fig. \ref{ED-PDH} is caused directly by the the
emergence of the sharp peak in $\Gamma({\bf k},\omega)$ shown in Fig. \ref{PDH-scattering-rate}.
Furthermore, we have also discussed the spectral line-shape in the electron-doped cuprate
superconductors in the normal-state, and found the sharp peak in $\Gamma({\bf k},\omega)$ that
can persist into the normal-state, indicating that the PDH structure is totally unrelated to
superconductivity, as the PDH structure in the hole-doped counterparts
\cite{Dessau91,Norman97,Campuzano99,Wei08,DMou17}.

\subsection{Dispersion kink}

The strong coupling of the electrons and various bosonic excitations in cuprate superconductors
manifested itself as a slope change (or a kink) in the quasiparticle dispersion
\cite{Damascelli03,Campuzano04,Fink07}. In the hole-doped cuprate superconductors, the quasiparticle
dispersion displays a kink at all around EFS \cite{Kaminski01,Lanzara01,Kordyuk06,Iwasawa08,Plumb13},
with the energy range $50\sim 80$ meV at which the kink appears. In spite of the extensive studies,
the controversy still exists on what bosonic mode causes the dispersion kink. In the recent studies
for the hole-doped case \cite{Liu20}, we have shown within the framework of the kinetic-energy driven
superconductivity that the quasiparticle dispersion is affected by a strongly dispersive spin
excitation, and then the dispersion kink associated with the dressing of the electrons is electronic
in nature. On the electron-doped side, although the experimental results of the PDH structure in the
quasiparticle excitation spectrum under the proper annealing condition are still lacking to date as
we have mentioned above in Section \ref{Introduction-1}, the dispersion kinks in the electron-doped
cuprate superconductor Pr$_{1.3-x}$La$_{0.7}$Ce$_{x}$CuO$_{4}$ under the proper annealing condition
along the nodal and antinodal directions have been detected very recently from the ARPES spectra
\cite{Horio20b}, where the {\it bare} band structure \cite{Horio20b,Horio18a} has been well fitted in
terms of the single-band tight-binding model with $t'/t=0.15\sim 0.19$. In this subsection, we discuss
the nature of the dispersion kink in the electron-doped cuprate superconductors, and adopt the
parameters $t/J=-2.5$ and $t'/t=0.16$ in the $t$-$J$ model (\ref{tjham}) for a quantitative or
semiquantitative comparison between the theory and experiment.

\begin{figure}[h!]
\centering
\includegraphics[scale=0.80]{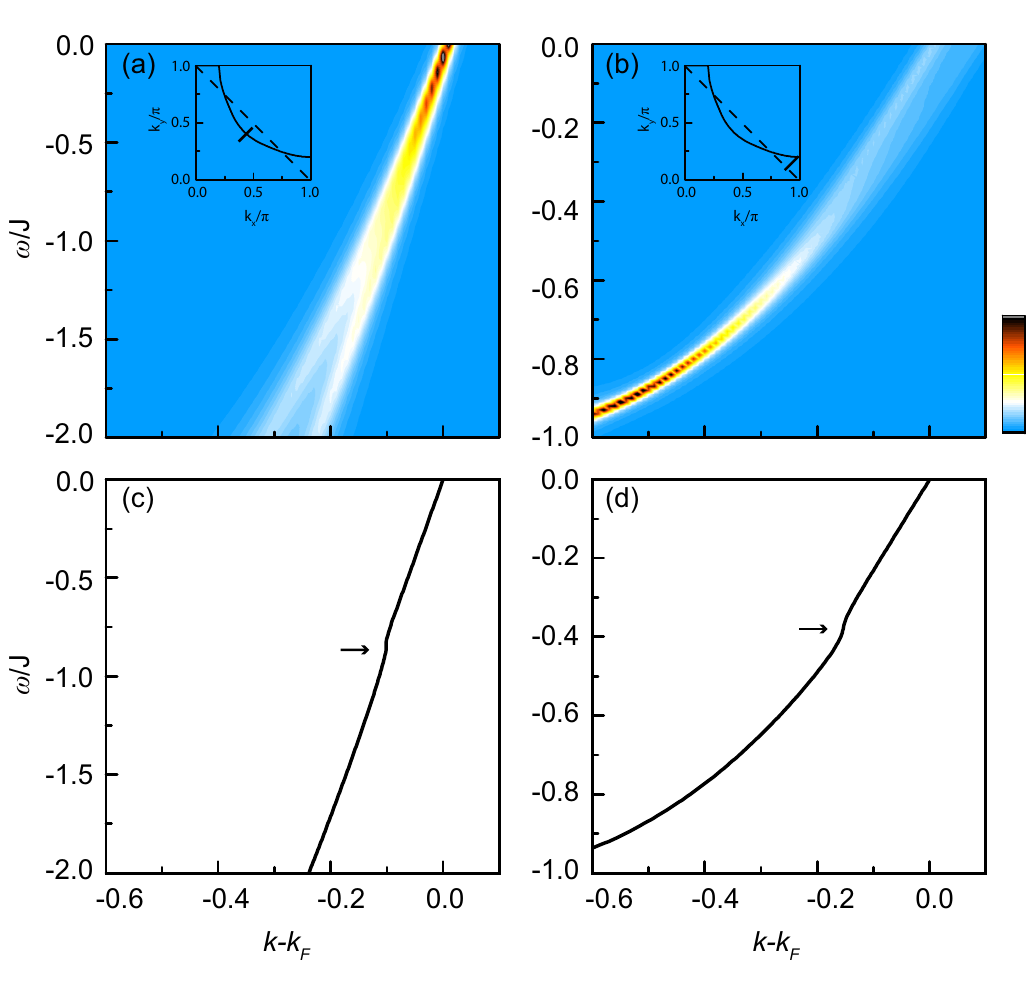}
\caption{(Color online) Upper panel: the intensity maps of the electron spectral function as a
function of binding-energy along (a) the nodal cut and (b) the antinodal cut at $\delta=0.15$
with $T=0.002J$ for $t/J=-2.5$ and $t'/t=0.16$, where the insets show the respective locations
of the two cuts in the Brillouin zone relative to the electron Fermi surface. Lower panel: the
quasiparticle dispersions along (c) the nodal cut and (d) the antinodal cut extracted from the
positions of the lowest-energy quasiparticle excitation peaks in (a) and (b), respectively,
where the arrow indicates the kink position. \label{kink-maps}}
\end{figure}

In Fig. \ref{kink-maps}, we plot the intensity map of the single-particle spectral function as a
function of binding-energy along (a) the nodal cut and (b) the antinodal cut at $\delta=0.15$ with
$T=0.002J$ for $t/J=-2.5$ and $t'/t=0.16$ in the upper panel . In the lower panel, we plot the
corresponding quasiparticle dispersions along (c) the nodal cut and (d) the antinodal cut extracted
from the positions of the lowest-energy quasiparticle excitation peaks in (a) and (b), respectively.
The arrow in Fig. \ref{kink-maps} marks the kink where the the linear quasiparticle
dispersion is separated as the low-energy and high-energy ranges with different slopes. It is
especially interesting to note that these results of the dispersion kink in the electron-doped
cuprate superconductors shown in Fig. \ref{kink-maps} are quite similar to those obtained in the
hole-doped counterparts \cite{Kaminski01,Lanzara01,Kordyuk06,Iwasawa08,Plumb13,Liu20}. In particular,
with a reasonably estimative value of $J\sim 100$ meV, the quasiparticle dispersion deviates from the
low-energy linear dispersion at around the kink energy $\omega_{\rm kink}\sim 0.79J=79$ meV along the
nodal cut, while this kink energy occurs at around $\omega_{\rm kink}\sim 0.40J=40$ meV along the
antinodal cut, which are semiquantitatively consistent with the corresponding results \cite{Horio20b}
observed on the electron-doped cuprate superconductor Pr$_{1.3-x}$La$_{0.7}$Ce$_{x}$CuO$_{4}$ under
the proper annealing condition along the nodal and antinodal directions, respectively. The above these
results in Fig. \ref{kink-maps} also indicate that although the kink in the quasiparticle dispersion
is present all around EFS, the kink energy decreases when the cut moves from the nodal region to the
antinodal region. Moreover, it should be noted that the coupling of the electrons with bosonic
excitations has been also observed early in all the electron-doped cuprate superconductors under the
improper annealing condition \cite{Santander-Syro09,Schmitt08,Park08a,Armitage03}, where the
quasiparticle dispersions present the kinks at around $50\sim 70$ meV along both the antinodal and
nodal cuts, which are also semiquantitatively consistent with our present results in
Fig. \ref{kink-maps}.

\begin{figure}[h!]
\centering
\includegraphics[scale=0.85]{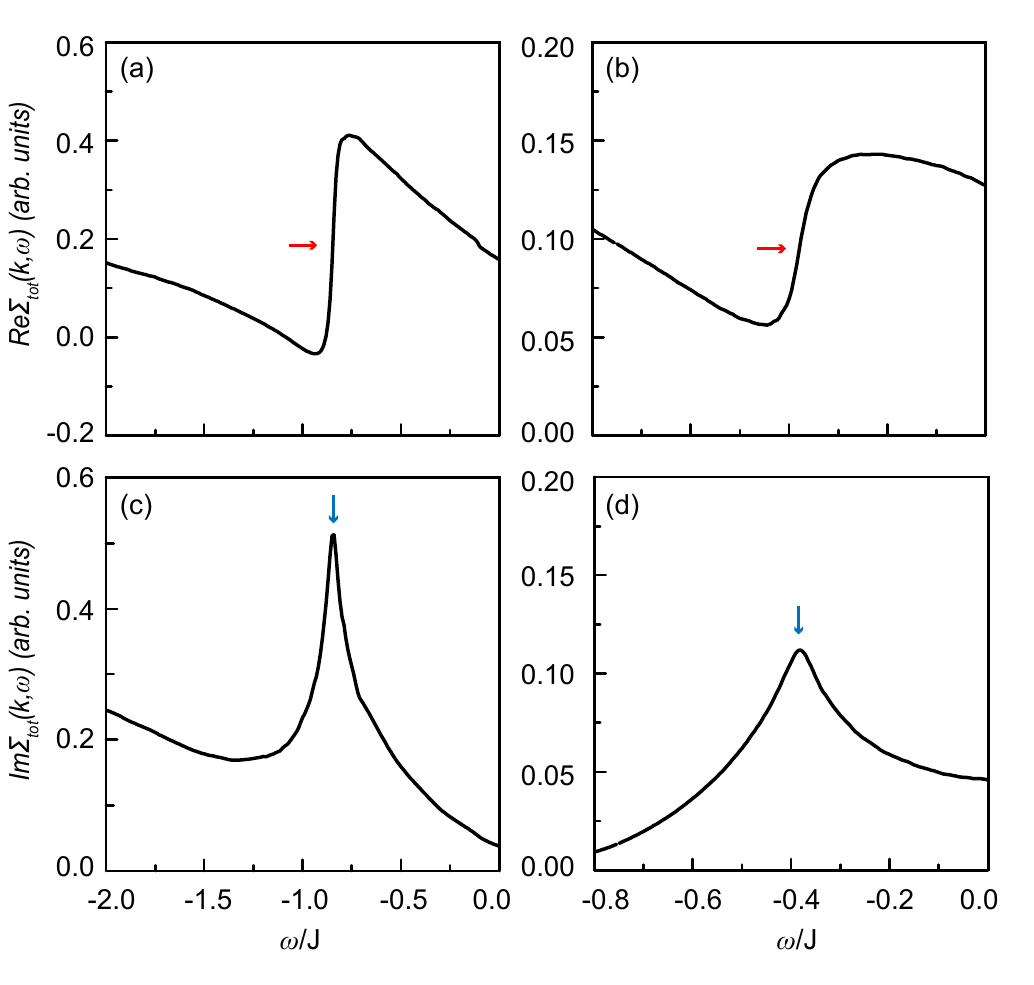}
\caption{(Color online) Upper panel: the real part of the total self-energy as a function of
binding-energy along (a) the nodal dispersion and (b) the antinodal dispersion at $\delta=0.15$
with $T=0.002J$ for $t/J=-2.5$ and $t'/t=0.16$, where the red arrow indicates inflection point.
Lower panel: the quasiparticle scattering rate as a function of binding-energy along (c) the
nodal dispersion and (d) the antinodal dispersion, where the blue arrow denotes the peak position.
\label{band-structure}}
\end{figure}

A natural and important question is which bosonic mode causes the dispersion kink in the
electron-doped cuprate superconductors? Within the framework of the kinetic-energy driven
superconductivity, the dispersion kink in the electron-doped cuprate superconductors arises from
the strong coupling between the electrons and a strongly dispersive spin excitation. This follows
a fact that from the quasiparticle dispersion in Eq. (\ref{band}), the dispersion kink shown in
Fig. \ref{kink-maps} does not originate from the single-electron band energy
$\varepsilon_{\bf k}$, but it is due to the slope change in the real part of the total
self-energy ${\rm Re}\Sigma_{\rm tot}({\bf k},\omega)$ , while the drop seen at around the
kink is directly associated with the corresponding peak in the quasiparticle scattering rate
$\Gamma({\bf k},\omega)$ in Eq. (\ref{rate}) [then the imaginary part of the total self-energy].
To further explore how the total self-energy generates the dispersion kink, we plot the real part
of the total self-energy ${\rm Re}\Sigma_{\rm tot}({\bf k},\omega)$ as a function of
binding-energy along (a) the nodal dispersion and (b) the antinodal dispersion as shown in
Fig. \ref{kink-maps}c and Fig. \ref{kink-maps}d, respectively, at $\delta=0.15$ with $T=0.002J$
for $t/J=-2.5$ and $t'/t=0.16$ in the upper panel of Fig. \ref{band-structure}, where the red
arrow indicates the inflection point (then the point of the slope change in the quasiparticle
dispersion). In the lower panel, we plot the corresponding quasiparticle scattering rate
$\Gamma({\bf k},\omega)$ as a function of binding-energy along (c) the nodal dispersion and (d)
the antinodal dispersion, where the blue arrow denotes the peak position (then the point of the
intensity depletion in the quasiparticle dispersion). It is obvious that a well-pronounced peak
in $\Gamma({\bf k},\omega)$ is clearly visible with a corresponding inflection point in
${\rm Re}\Sigma_{\rm tot}({\bf k},\omega)$ that appears at the exactly same energy in
 $\Gamma({\bf k},\omega)$. This sharp peak in $\Gamma({\bf k},\omega)$ suppresses heavily the
spectral weight at around the inflection point. More importantly, we find that the position of
the dispersion kink in Fig. \ref{kink-maps} matches exactly with the position of the inflection
point in ${\rm Re}\Sigma_{\rm tot}({\bf k},\omega)$ [then the position of the peak
in $\Gamma({\bf k},\omega)$] in Fig. \ref{band-structure}. In other words, the emergence of the
dispersion kink is determined by both the inflection point in
${\rm Re}\Sigma_{\rm tot}({\bf k},\omega)$ and the well-pronounced peak in
$\Gamma({\bf k},\omega)$. This is why the dispersion kink marks the crossover between two
different slopes, while the weak spectral intensity appears always at around the kink position
\cite{Santander-Syro09,Schmitt08,Park08a,Armitage03}.

\subsection{Peak-structure in self-energy}

In ARPES experiments \cite{Damascelli03,Campuzano04,Fink07}, the energy distribution curves are
observed when the photoemission intensity is measured at constant momentum, while the momentum
distribution curves are observed when the photoemission intensity is measured at constant energy.
The information of the energy and momentum dependence of the electron self-energy then can be
extracted in terms of the single-particle spectral function (\ref{ESF}) by analyzing the spectral
intensity of the energy and momentum distribution curves. However, as shown in the single-particle
spectral function (\ref{ESF}), only the total self-energy can be extracted directly from ARPES
experiments \cite{Damascelli03,Campuzano04,Fink07}. In particular, for our exploration of the
strongly dispersive spin excitation coupling in the kinetic-energy driven SC mechanism that is how
the normal and anomalous self-energy effects compete in the SC-state, it thus is crucial to extract
the normal and the anomalous self-energies separately. This also follows a basic fact that in the
kinetic-energy driven superconductivity \cite{Feng15a,Feng0306,Feng12,Feng15}, the SC-state is
controlled by both the electron pair gap and single-particle coherence, where the single-particle
coherence is dominated by the normal self-energy $\Sigma_{\rm ph}({\bf k},\omega)$, and therefore
antagonizes superconductivity, while the energy and momentum dependent electron pair gap is
determined completely by the anomalous self-energy $\Sigma_{\rm pp}({\bf k},\omega)$, and
therefore is corresponding to the energy for breaking an electron pair. In this case, if both the
normal and anomalous self-energies are deduced from the experimental data, it can be used to
examine the microscopic theory of the kinetic-energy driven superconductivity and understand the
essential physics of cuprate superconductors.

In the hole-doped cuprate superconductors, the machine learning technique has been
applied recently to deduce both the normal and anomalous self-energies from the experimental data
of the ARPES spectra observed in Bi$_{2}$Sr$_{2}$CaCu$_{2}$O$_{8+\delta}$ at the optimum doping
and Bi$_{2}$Sr$_{2}$CuO$_{6+\delta}$ in the underdoped regime \cite{Yamaji19}, and the deduced
results show clearly that both the normal and anomalous self-energies exhibit the sharp low-energy
peak-structures. Moreover, these machine learning 'experimental' results confirm that the peak in
the anomalous self-energy makes a dominant contribution to the SC gap, and therefore provide a
decisive testimony for the origin of superconductivity \cite{Yamaji19}. In the very recent studies
\cite{Liu20a}, we have made a comparison of these deduced normal and anomalous self-energies in
Ref. \onlinecite{Yamaji19} with those obtained based on the kinetic-energy driven SC mechanism,
and then show explicitly that the interaction between electrons by the exchange of spin
excitations generates the sharp low-energy peak-structures both in the normal and anomalous
self-energies, in striking agreement with the corresponding results in the normal and anomalous
self-energies deduced via machine learning 'experiments' \cite{Yamaji19}. In this subsection, we
analyze the normal and anomalous self-energies in the electron-doped cuprate superconductors, and
then compare these normal and anomalous self-energies with those obtained in the hole-doped
counterparts \cite{Yamaji19,Liu20a}.

\begin{figure}[h!]
\centering
\includegraphics[scale=0.80]{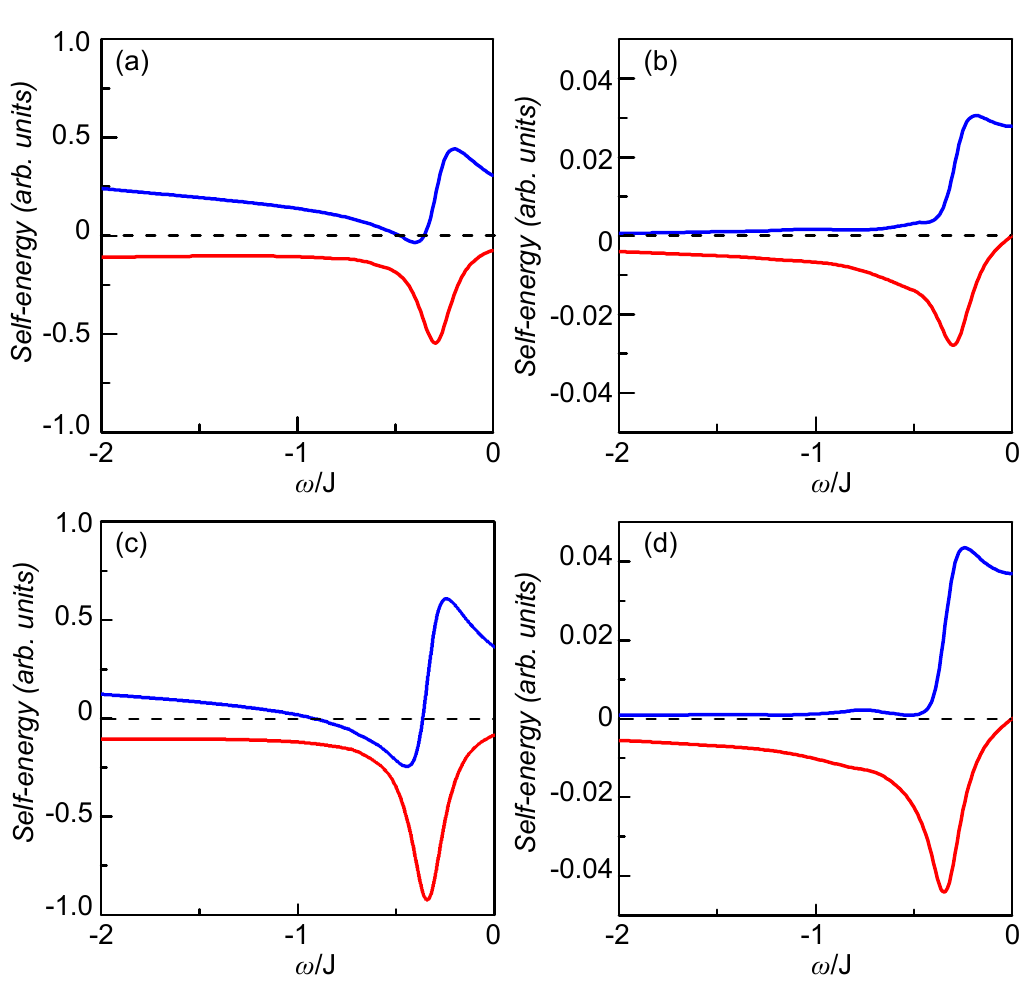}
\caption{(Color online) (a) The real (blue line) and imaginary (red line) parts of the normal
self-energy and (b) the real (blue line) and imaginary (red line) parts of the anomalous
self-energy at the antinode as a function of energy in the electron doping $\delta=0.15$ with
$T=0.002J$ for $t/J=-2.5$ and $t'/t=0.3$. The corresponding results of (c) the real and
imaginary parts of the normal self-energy and (d) the real and imaginary parts of the anomalous
self-energy of the hole-doped cuprate superconductors at the antinode as a function of energy
in the hole doping $\delta=0.15$ with $T=0.002J$ for $t/J=2.5$ and $t'/t=0.3$ taken from
Ref. \onlinecite{Liu20a}. \label{self-energy-AN}}
\end{figure}

In Fig. \ref{self-energy-AN}, we plot (a) the real (blue line) and imaginary (red line) parts of
the normal self-energy and (b) the real (blue line) and imaginary (red line) parts of the
anomalous self-energy at the antinode as a function of energy in the electron doping $\delta=0.15$
with $T=0.002J$ for $t/J=-2.5$ and $t'/t=0.3$. For a better comparison, the corresponding results
\cite{Liu20a} of (c) the real and imaginary parts of the normal self-energy and (d) the real and
imaginary parts of the anomalous self-energy at the antinode as a function of energy in the
{\it hole} doping $\delta=0.15$ with $T=0.002J$ for $t/J=2.5$ and $t'/t=0.3$ are also shown in
Fig. \ref{self-energy-AN}. Surprisedly, both the normal and anomalous self-energies in the
electron-doped cuprate superconductors also exhibit the sharp low-energy peak-structures. More
specifically, the main features of these low-energy peak-structures are in a stark similarity with
the corresponding low-energy peak-structures in the normal and anomalous self-energies of the
hole-doped cuprate counterparts obtained based on the kinetic energy driven superconductivity
\cite{Liu20a}, and deduced from the ARPES spectra via machine learning \cite{Yamaji19}. This is
why the absence of the disparity between the phase diagrams of the electron- and hole-doped
cuprate superconductors can be observed experimentally \cite{Song17,Lin20,Horio20b}. In the
SC-state, the pairing force and electron pair order parameter have been incorporated into the
anomalous self-energy $\Sigma_{\rm pp}({\bf k},\omega)$, which is identified as the
SC gap, and therefore describes the strength of the electron pair. In this case, the dominant
contribution to the strength of the electron pair arises from the sharp low-energy peaks both in
${\rm Re}\Sigma_{\rm pp}({\bf k}_{\rm AN},\omega)$ and
${\rm Im}\Sigma_{\rm pp}({\bf k}_{\rm AN},\omega)$ \cite{Yamaji19,Liu20a}. On the other hand, from
the total self-energy in Eq. (\ref{TOT-SE}), the dressing of the electrons are mainly dominated by
the normal self-energy $\Sigma_{\rm ph}({\bf k}_{\rm AN},\omega)$, indicating that the sharp
low-energy peak in ${\rm Re}\Sigma_{\rm ph}({\bf k}_{\rm AN},\omega)$ offsets mainly the
single-electron band energy, while the sharp low-energy peak in
${\rm Im}\Sigma_{\rm ph}({\bf k}_{\rm AN},\omega)$ governs mainly the lifetime of the quasiparticle
excitation \cite{Liu20a}. However, the sharp low-energy peak in the anomalous self-energy
disappears in the normal-state, while the sharp low-energy peak in the normal self-energy can
persist into the normal-state. This is also why the exotic features, including the intrinsic
EFS reconstruction, the PDH structure in the quasiparticle excitation spectrum, the dispersion kink,
and the ARPES autocorrelation and its connection with the quasiparticle scattering interference
\cite{Tan20}, arisen from the dressing of the electrons that emerge in the SC-state also appear
in the normal-state.

For a further understanding of the essential physics of the electron-doped cuprate superconductors,
we have also studied the doping and momentum dependence of the low-energy peak structures both in
the normal and anomalous self-energies, and all the obtained results are consistent with the
corresponding results obtained in the hole-doped counterparts \cite{Yamaji19,Liu20a}. We therefore
complete the picture of the dressing of the electrons for the electron-doped cuprate superconductors,
and show the existence of the common origin of the anisotropic dressing of the electrons both in the
electron- and hole-doped cuprate superconductors, i.e., the dressing of the electrons arises from
the interaction of the electrons with a strongly dispersive spin excitation. Since the remarkable
low-energy peak-structures both in the normal and anomalous self-energies of the hole-doped cuprate
superconductors \cite{Liu20a} are well consistent with those deduced via machine learning
'experiments' \cite{Yamaji19}, we therefore predict that the similar low-energy peak-structures both
in the normal and anomalous self-energies can be also deduced from the experimental data of the ARPES
spectra observed on the electron-doped cuprate superconductors in the proper annealing condition
via machine learning.

\section{Conclusions}\label{conclusions}

Within the framework of the kinetic-energy driven superconductivity, we have studied the phase diagram
of the electron-doped cuprate superconductors and the related exotic features of the anisotropic
dressing of the electrons. Our results show that although the optimized $T_{\rm c}$ in the
electron-doped superconductors is much smaller than that in the hole-doped counterparts, the electron-
and hole-doped cuprate superconductors rather resemble each other in the doping range of the SC dome,
indicating an absence of the disparity between the phase diagrams of the electron- and hole-doped
cuprate superconductors. In particular, the anisotropic dressing of the electrons due to the strong
electron's coupling to a strongly dispersive spin excitation induces a EFS reconstruction, where the
closed EFS contour in the case without considering the strong electron interaction is broken up into
the disconnected Fermi arcs centered around the nodal region, in qualitative agreement with the
available experimental data. Moreover, we reveal the spin excitation effect in the quasiparticle
excitation spectrum, where the dip in the PDH structure developed in the quasiparticle excitation
spectrum at around the antinodal (nodal) region is directly related to the corresponding sharp
peak in the antinodal (nodal) quasiparticle scattering rate, while the appearance of the dispersion
kink is always associated with the emergence of the inflection point in the real part of the total
self-energy and the sharp peak in the quasiparticle scattering rate. By comparing directly with the
corresponding results in the hole-doped cuprate superconductors, our present results therefore show
that the EFS reconstruction to form the Fermi arcs, the PDH structure in the quasiparticle excitation
spectrum, and the dispersion kink, are the intrinsic properties both of the electron- and hole-doped
cuprate superconductors. Although some subtle differences between the electron- and hole-doped
cuprate superconductors appear due to the electron-hole asymmetry, the coupling of the electrons with
a strongly dispersive spin excitation is the common origin for the anisotropic dressing of the
electrons both in the electron- and hole-doped cuprate superconductors. The theory also predicts that
both the normal and anomalous self-energies exhibit the notable low-energy peak-structures, which
should be verified by future machine learning 'experiments'.

~~\\
\section*{Acknowledgements}

This work was supported by the National Key Research and Development Program of China under
Grant No. 2016YFA0300304, and the National Natural Science Foundation of China under Grant
Nos. 11974051 and 11734002.

\end{document}